# Understanding Quantum Theory
Michael Drieschner

*I think I can safely say that nobody understands quantum mechanics.*
*Richard Feynman*[1]

## Inhaltsverzeichnis

## 1 Introduction

It is now about 100 years ago that Erwin Schrödinger[2] published his quantum theory, a few months after Heisenberg[3] had published his. Very soon Schrödinger gave a proof that both formalisms are physically equivalent. Thus, for almost 100 years now the formalism of quantum mechanics seems clear. For the same time there has been a dispute, however, about the interpretation of that formalism, which is not settled yet even after 100 years. How is that possible, after those 100 years of extremely successfully applying quantum mechanics in physics as well as in technology?

Why is it that a physical formalism needs an interpretation at all? In classical mechanics it seems entirely clear how the results of the formalism have to be connected with empirical findings. Nobody ever called especially for an "interpretation" of classical mechanics. But for quantum mechanics that is different: The result of a quantum mechanical calculation is usu-ally not the value of an observable but typically a probability or a probability density. So the question arises what that means from the viewpoint of physics—it calls for an interpretation.

---

[1] Richard Feynman, *The Character of Physical Law* (1965). Transcript of the Messenger Lectures at Cornell University, presented in November 1964. London etc. (Penguin) 1992. Chapter 6, "Probability and Uncertainty — the Quantum Mechanical View of Nature," p. 129
[2] Schrödinger, E.: 1926. Ann.Phys. **79**, 361, 489, 734; **80**, 437; **81**, 109
[3] Heisenberg, W.: 1925: Zeitschrift für Physik **33**, p. 879

What I attempt here is making understandable what quantum theory really says. It will be different from most well-known "popular" accounts of quantum theory, in that I do not want to present you that "miraculous" modern physics. Many popular descriptions of quantum theory concentrate on presenting a kind of curiosity cabinet with news like "a particle can be a wave as well" or "a particle might be at different places at the same time" or "quantum theory allows instant transfer of information over large distances". Curios like that distort what quantum theory really says, and they reliably prevent any understanding of that theory. I will try to clarify, instead, what quantum theory is about and what it says—in rather sober words that give really a better "understanding".

Thus, I am doing something similar to what Leonard Susskind and Art Friedman do in their '*Quantum Mechanics. The Theoretical Minimum.*'[4] Their excellent account introduces into *physical* essentials of quantum theory whereas this text attempts to give an introduction on how to *understand* quantum theory, contrary to Feynman's quote in the beginning. Whether Feynman is right depends, evidently, on what we understand by understanding. This is actually a philosophical question. The answer might become clearer in the course of this text.

## 2 Quantum vs. "Classical" Ontology

Classical ontology assumes that there is a nature "out there" we can watch and describe like one would describe e.g. the works of a clock. This view is described classically by P.S. Laplace[5]:

"Une intelligence qui, pour un instant donné, connaîtrait toutes les forces dont la nature est animée, et la situation respective des êtres qui la composent, si d'ailleurs elle était assez vaste pour soumettre ces données à l'analyse, embrasserait dans la même formule, les mouvements des plus grand corps de l'univers et ceux du plus léger atome : rien ne serait incertain pour elle, et l'avenir comme le passé, serait présent à ses yeux."[6]

If you look close enough you can see that quantum theory is different from that picture. That is what I think leads many people to finding quantum theory so strange. What you can get from quantum theory is a system of predictions for possible measurements, and those predictions do not generally give certainty to an outcome but admit different possible outcomes with respective probabilities. It is true, probabilities occur in classical physics as well. But there one can always comfort oneself with the idea that "in reality" one outcome was certain, only the information the physicist had was not sufficient to obtain that "real" outcome. In quantum theory, this way out is barred. No quantum state that describes a system is such that it gives definite values to all observables. It rather gives for most observables predictions and, in the optimal case, for every possible result a probability for finding that result. Thus, we have to admit that quantum theory is a fundamentally indeterministic theory. That insight forces us to abandon the "classical" picture of a world "out there" comparable to a clock work.

---

[4] Leonard Susskind and Art Friedman, Quantum Mechanics. The Theoretical Minimum. London (penguin) 2015
[5] Pierre Simon de Laplace, *Essai philosophique sur les probabilités*. Paris 1814
[6] English version [MD]: "An intelligence which, in a given instant, knew all forces that animate nature and the correlations of the beings it is made up of; if it were, besides, huge enough to analyze these data, it would comprise in the same formula the movement of the largest bodies of the universe as well as those of the lightest atom: Nothing would be uncertain for it, and the future like the past were present before its eyes."

Is that a drawback, compared to the "classical" ontology? We have to reflect on what we really expect of science: science should enable us to predict what will happen, using empirical data about the present state and theoretical calculations on the basis of a valid theory. In classical physics we had the singular situation that we could predict e.g. the positions of a planet, in principle, with certainty. This was, however, very principle-like; it applied mainly to the standard case of the movement of a planet. Even there, it is generally impossible to know the present situation and all influences on it exactly enough to predict with certainty. Thus, in the "classical" situation probabilities come in as well, not different from quantum theory. But still, in the foundations there is that decisive difference between a deterministic and an indeterministic theory: in a deterministic, "classical" theory we can always suppose that in fact every observable has an exact value.

When quantum theory was discovered, it was not clear from the beginning that indeterminism was its key feature. It was only after many futile attempts that Max Born published his seminal paper (1926)[7]. There he says that the square of the coefficient in the decomposition of the wave function is the probability of finding the corresponding result. Born continues his considerations with the suggestion that quantum theory is fundamentally indeterministic.

Among the "Copenhagen" experts, there had been speculations about indeterminism for some time before but most of them hesitated publishing their opinion because of the strong impact that such a change must have on our picture of the world. Thus Born was the first one who dared publishing that consequence of their discussions.

## 3 Indeterminism

In fact, indeterminism is the revolutionary new property of quantum theory, compared with all other physical theories. Indeterminism is the one property of quantum theory that is, to my mind, the real reason for the difficulties felt with understanding that theory.

Now, indeterminism does not mean that there is no way of predicting anything about the result of future measurements. Instead of predictions with certainty we have, in an indeterministic theory, predictions with probability. As a consequence, many difficulties felt with quantum theory can be reduced to problems of probability. Thus, our first question is now, what is probability?

## 4 Probability[8]

The probability of a result, as used in empirical science, has to do with the relative frequency of that result in a series of measurements. Many physicists simply identify probability with relative frequency. But that does not work: Imagine throwing a coin, with probability of heads and tails both being ½. Now flip that coin 13 times. The relative frequency ½ would mean 6.5 results heads. That is not possible! – But there is still another objection that goes deeper: Using probability calculus we can calculate—we shall deal with that later—probabilities on a higher level. Consider, as an example, a series of 12 tosses of dice, and repeat that series many times. The probability, e.g., with a "good" die to get a 'four' is 1/6. From probability calculus we can calculate the probability that *exactly* 1/6 of the 12, i.e. 2 throws will show a 'four'. The formula for the probability $p(n)$ of a number of $n$ results 'four' in the sequence of 12 throws ($n = 0, \ldots, 12$) is:

$$p(n) = \binom{12}{n} \cdot \left(\frac{1}{6}\right)^n \cdot \left(\frac{5}{6}\right)^{12-n}.$$

---

[7] Max Born, 'Zur Quantenmechanik der Stoßvorgänge.' Zeitschrift f. Physik 37, 863–867. here p. 865
[8] https://plato.stanford.edu/archives/sum2003/entries/probability-interpret/

The following table gives values $p(n)$:

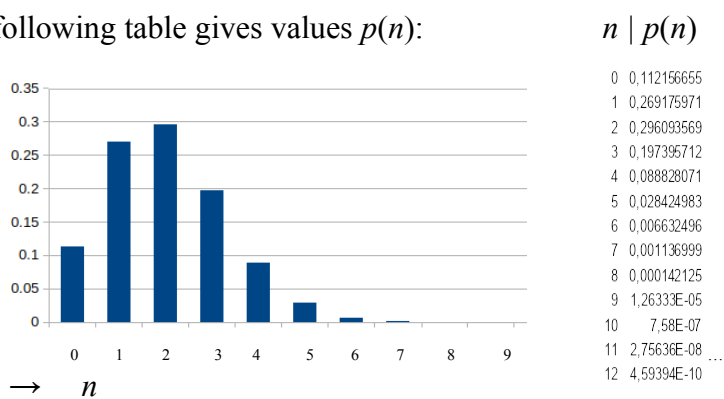

| $n$ | $p(n)$ |
|---|---|
| 0 | 0,112156655 |
| 1 | 0,269175971 |
| 2 | 0,296093569 |
| 3 | 0,197395712 |
| 4 | 0,088828071 |
| 5 | 0,028424983 |
| 6 | 0,006632496 |
| 7 | 0,001136999 |
| 8 | 0,000142125 |
| 9 | 1,26333E-05 |
| 10 | 7,58E-07 |
| 11 | 2,75636E-08 |
| 12 | 4,59394E-10 |

This gives $p(2) = 29.6\%$, but the probability to get only one "four" is not much less, namely $p(1) = 26.9\%$, and, just giving a few more examples, $p(3) = 19.7\%$, $p(0) = 11.2\%$, and $p(12) = 5 \cdot 10^{-10}$. That means: *All* relative frequencies, not only 1/6, have positive probability, i.e. they are possible, and relative frequencies unequal but near the value of the probability are almost as probable as those exactly equal to the probability. Thus, probability theory itself excludes the *identification* of probability with relative frequency. We will come back later to such considerations.

### *Predictions*

It can be shown that "predicted relative frequency" is a good definition of probability. From this definition even the rules of probability calculus can be deduced.[9]

## 5 The Necessity of Classical Concepts

There is a great problem with an indeterministic theory one would not think of in the beginning: How can we use such a theory for describing reality? – Quantum mechanics gives predictions of the form: "When quantity $Q$ is measured then one will find the value $q_1$ with probability $p(q_1)$." But that gives no possibility for describing *reality*, quantum mechanics does not give a picture of a reality "out there"! This difference between quantum mechanics and classical physics seems to be the reason of the uneasiness some people felt immediately after the discovery of quantum mechanics, and some still feel uneasy today.

There is one area where that lack is especially felt, namely in the description of measurements. Niels Bohr remarked that we have to be able to say "…what we have done and what we have learnt." His conclusion for quantum mechanics is that classical concepts are necessary for a description of measurements. In classical theory it is presupposed that descriptions like "The present apparatus measures quantity $Q$", or "The measured value is $q_1$" make sense. Since within the framework of quantum mechanics such descriptions of reality do not occur, we have to take the aid of classical physics in order to be able to give such descriptions. But there we have a problem: According to quantum mechanics classical physics is outdated; wherever quantum mechanics and classical physics give different results, quantum mechanics is right, classical physics is wrong. – This sounds awful, but actually there is no reason for worrying, because really in the usual cases quantum mechanics gives *almost* the same results as classical physics. A working physicist, in any case, would not consider that a catastrophe since in physics, as mentioned above, we rarely have exact values at all. Still, many philosophers of science, who traditionally come from logic or mathematics, would insist that "almost correct" is, in fact, false. But in physics that argument is not valid. Approximation lies at the foundation of physics; without approximation physics is not possible at all. Maybe the most important argument is: Physics deals with objects which are defined by the observables that can be measured on those objects. Take, e.g., a classical point mass, which is defined as an object the state of which at a certain instant can be described completely by its position and momentum. There are no things in the world that are point masses. In order to describe anything as a point mass we use approximations: We discard every information except about position and momentum; we treat the mass of the body concerned as concentrated in one point in space; we give the results of our measurements real number values—all of which do not depict reality as it is but in a certain approximation. Without that we could not do physics.

---

[9] cf. Michael Drieschner, Found. Phys. 46(2016)28–43

It is actually quite normal to accept a proposition being 'approximately true' as a true proposition. Thus in physics there is no problem in accepting a "classical" proposition within the framework of quantum mechanics. And more than that: quantum mechanics is incomplete without those parts of classical physics; it could not be linked to reality without.

## 6 Interpretations[10]

In the course of almost 100 years of discussions about quantum theory, so many interpretations have been proposed that I will not endeavor giving a survey, and much less giving a systematic account of that abundance. My purpose here is rather to give some examples that may make my description clearer. Therefore, I will give short comments on three very different interpretations, namely
- the "Bohmian mechanics" introduced by David Bohm in 1952[11];
- the "many-worlds" interpretation introduced by Everett, Wheeler, and DeWitt from 1957 on[12];
- and last, but not least, the "Minimal Interpretation of Quantum Mechanics" I will present here.

I did not mention the "Copenhagen Interpretation". Since it is universally acknowledged that it is difficult to get a clear cut definition of what that interpretation is, I rather do not attempt to give one. Possibly the "Minimal Interpretation" presented here comes close to what is usually called the Copenhagen Interpretation.

---

[10] cf. Lewis, Peter J., "Interpretations of Quantum Mechanics", *Internet Encyclopedia of Philosophy*.
[11] Bohm, David: Phys.Rev. **85** (1952) 166,180; reprinted in: Wheeler, J.A. and Zurek, W.H. (eds.): Quantum Theory and Measurement. Princeton, NJ (UP) 1983. There is a very well informed and thorough account in: Passon, Oliver, Bohmsche Mechanik. Frankfurt/M (Harri Deutsch) 2004. (there is no English translation).
[12] DeWitt, Bryce S.; Graham, R. Neill (eds.): *The Many Worlds Interpretation of Quantum Mechanics.* Princeton University Press, 1973.

## *Minimal Interpretation*

In March 2000 there was in the journal 'Physics Today' an article "Quantum Theory needs no 'interpretation'" which stated, among others: "The thread common to all the non-standard "interpretations" is the desire to create a new theory with features that correspond to some reality independent of our potential experiments. But, trying to fulfill a classical worldview by encumbering quantum mechanics with hidden variables, multiple worlds, consistency rules, or spontaneous collapse, without any improvement in its predictive power, only gives the illusion of a better understanding. Contrary to those desires, quantum theory does not describe physical reality. What it does is provide an algorithm for computing probabilities for the macroscopic events ("detector clicks") that are the consequences of our experimental interventions. This strict definition of the scope of quantum theory is the only interpretation ever needed, whether by experimenters or theorists."[13]

Our concept of 'Minimal Interpretation' appears, as well, in a modern overview book[14]. The authors (in this case, Cord Friebe) state:"If one tries to proceed systematically, then it is expedient to begin with an interpretation upon which everyone can agree, that is with an instrumentalist minimal interpretation. In such an interpretation, Hermitian operators represent macroscopic measurement apparatus, and their eigenvalues indicate the measurement outcomes (pointer positions) which can be observed, while inner products give the probabilities of obtaining particular measured values. With such a formulation, quantum mechanics remains stuck in the macroscopic world and avoids any sort of ontological statement about the (microscopic) quantum-physical system itself." Thus, we have here the contrary of the none-interpretation mentioned above. They continue their account with: "Going one step further, we come to the ensemble interpretation: Here, the mathematical symbols indeed refer to microscopic objects, but only to a very large number of such systems. According to this view, quantum mechanics is a kind of statistical theory whose laws are those of large numbers. In regard to a particular system, this interpretation remains agnostic." – It is interesting that in this last interpretation probability is introduced as "laws of large numbers". This is a very special interpretation of probability and does not directly concern quantum mechanics at all.

Friebe et al. do not seem to be satisfied with the "minimal interpretation" described above. They state (p. 44):"The fact that this minimal interpretation makes statements only about macroscopic, empirically directly accessible entities such as measurement setups, particle tracks in detectors or pulses from a microchannel plate may be quite adequate for those who see the goal of the theory within an experimental science such as physics as being simply the ability to provide empirically testable predictions. For the metaphysics of science, this is not sufficient, and most physicists would also prefer to have some idea of what is behind those measurements and observational data, i.e. just how the microscopic world which produces such effects is really structured." Continuing, they are certainly right in suspecting: "In contrast to the instrumentalist minimal interpretation, however, every additional assumption which might lead to a further-reaching interpretation remains controversial."

---

[13] Fuchs, Christopher A.; Asher Peres: "Quantum Theory needs no 'interpretation'". Physics Today, March 2000, p. 70-71; here p.70

[14] Friebe (2018)

Here you find the decisive keys: "what is behind?" and "how the microscopic world [...]is really structured." Those who ask these questions apparently presuppose that *there is* something behind, that *there is* a microscopic world that is *really structured* somehow. But, how do they know? Is there some necessity for such presuppositions? – To my mind it is just our being accustomed to the classical ontology, as I called it above, that leads us to believing that. Actually quantum mechanics seems rather to show that such questions lead nowhere. We shall see in the discussion further on that the alleged solutions of this task offer nothing but an additions of words to the well-known theory—one should rather do without.

There is, though, a good sense in attributing properties to objects of quantum mechanics: If it is possible to predict with certainty the outcome $x_o$ for a measurement of the quantity **X** on object[15] **O**, we can say (as a kind of abbreviation) "The object **O** *has* property $x_o$".

What we called here "Minimal Interpretation" is a kind of replacement for the rather foggy picture we have of the Copenhagen interpretation. This "Minimal Interpretation" is what I think I can understand and justify as a convincing interpretation of quantum mechanics, and in some way it resembles Copenhagen interpretations.

## 7 Mathematical formalism of quantum mechanics

For the further discussion, let us start from the mathematical formalism: In quantum mechanics a physical object is described with the help of a complex vector space with an inner product ("Hilbert space"). How is physics connected with that space? – The observables of the object described are represented by self-adjoint operators of the Hilbert space. An eigenvalue $x_0$ of a self-adjoint operator *X* represents a possible result of a measurements of the corresponding observable *X*. The present state of the object is described by a vector of norm 1 in the vector space (more abstractly, it is described by the eigenspace of the said eigenvalue; if the eigenspace is one-dimensional, it is conventionally represented by a vector in it of norm one. Still more generally, the state of the system is represented by a "statistical operator"). The eigenvector $\xi_0$ corresponding (in the special case mentioned) to the eigenvalue $x_0$ represents the state of the system after the eigenvalue $x_0$ has been measured. If the state of the system before the measurement is correctly described by the state-vector $\xi$ (of norm 1) then the probability of measuring value $x_0$ is the square of the absolute value of the inner product of the two vectors,

$$p(x_0) = |(\xi,\xi_0)|^2,$$

provided $(\xi_0,\xi_0) = (\xi,\xi) = 1$ (ignoring for the time being a more general description).

The time development of the state $\xi$ of the system is described by a unitary transformation depending on to the elapsed time *t*,

$$\xi(t) = e^{\frac{-i}{\hbar} \cdot H \cdot t} \cdot \xi(0)$$

where *H* is the Hamiltonian Operator, which represents the observable 'energy'.

## 8 Lattice

The above presentation describes quantum mechanics as students learn it. It leads to some specific difficulties, though, for understanding quantum mechanics. One could, e.g., ask what the addition of vectors of Hilbert space means in reality—since it occurs in the mathematical description that is supposed to represent a corresponding structure of reality. But asking such a question does not make sense: 'addition of vectors in Hilbert space' does not have a meaning in a description of reality. This is important to know for an interpretation: The topic of admitting questions arises only from a formulation of quantum mechanics that does not easily lend itself for an understanding, namely in so far as it uses the vector space formalism.

---
[15] We treat the terms 'object' and 'system' as synonyms

This drawback might be cured by a more abstract but more comprehensible mathematical picture in using the concept of *lattice*:

A mathematical *lattice* is a partially ordered set (with an ordering we signify by '$\leqslant$') which is closed under meet ($\cap$) and join ($\cup$), where *meet*, in this general context, is the greatest lower bound according to the order relation '$\leqslant$', and *join* is the least upper bound. A lattice contains the elements $\emptyset$ and $\mathbf{1}$, the *minimal* and *maximal* elements of the whole structure, according to the order relation '$\leqslant$'. A special example of a lattice is the set of all subsets of a given set, ordered by set inclusion ($\subseteq$)—isomorphic with the lattice of (classical) propositional logic. This type of lattice is called Boolean. But the concept of lattice is much more general.

Let us be more specific for our case of quantum mechanics: A lattice is *orthocomplemented* iff for every element $E$ of the lattice there is an *orthocomplement,* namely an element $E^\perp$ such that

- $E \cap E^\perp = \emptyset$;
- $E \cup E^\perp = \mathbf{1}$;
- $E^{\perp\perp} = E$;
- if $E \leqslant F$ then $F^\perp \leqslant E^\perp$.

Again, the set of subsets of a given set, ordered by set inclusion ($\subseteq$), is a special example, where the orthocomplement is the ordinary set complement. An example that is important for quantum mechanics is the set of all subspaces of Hilbert space. It is ordered by set inclusion, the orthocomplement of a subspace is the orthogonal subspace.

Let us now consider the structure of all properties of a given physical system or, maybe more realistic, the structure of all possible results of measurements of that system! The ordering relation in that case is *implication* ('$\to$'): $A \to B$ means: 'Whenever $A$ is necessary then $B$ is necessary as well'. The other properties of the orthocomplemented lattice are the the following: the orthocomplement is *negation* ('not', $\neg$); the 'meet' operator represents *conjunction* ('and', $\wedge$)—nothing but putting two predictions beside each other. From there, *disjunction* ('or', $\vee$) can be derived as:
$$A \vee B = (A^\perp \wedge B^\perp)^\perp.$$

It can easily be seen from the definition of the orthocomplement that $A \wedge B$ is the greatest lower bound, and $A \vee B$ is the least upper bound of $A$ and $B$.

Again, classical propositional logic is an example, as well as the lattice of subspaces of Hilbert space. Because evidently there are similarities between the two, the lattice of the subspaces of a Hilbert space sometimes is called "Quantum Logic"[16].

Considering quantum mechanics from the lattice point of view gives us the means at hand that make it easier to understand quantum mechanics—without being obliged to dive into the depths of differential analysis or the like. All we have to do is fixing a few fundamentals of what we understand physics to be.

# 9 Physical objects ("systems")

The first question: What is physics about? – One possible answer is: "Physics gives the opportunity for predictions, based on the present state and using established theories." – This answer is felt by some scholars as being too "operational."[17] They are rather seeking an objective description of reality. But these two goals of physics are not as far apart as it might seem. What is an 'objective' description? It is one that is valid independently of the person, time and place; it can be verified—in principle—by anyone anytime anywhere. 'Verifying' means, checking whether the respective proposition is true. This kind of checking is possible only if the proposition is a *prediction* which can come true or not. Thus, any objective description necessarily involves predictions.

Thus, we can record that physics deals with objective predictions. – How are such predictions connected among each other? – Actually, as mentioned above, the predictions do not concern the real world but an approximate, idealized "model", the *physical system*. It is defined by its observables. Here the fundamental role of predictions comes into play: A physical system is composed of those observables that make, together, predictions about the same observables possible. Take, e.g., the classical object 'point mass'. Its defining observables are position and momentum. Momentum governs the change of position, and how momentum will change (i.e. the forces on the point mass) depends in many cases only on the position and, maybe in addition, on momentum (as for friction). Thus, for many cases position and momentum are a good choice of observables for a system to enable predictions about those same observables. Therefore, a point mass is a good object in such cases. – Let us have a look at a more complex example, the electromagnetic field, defined by its values at every point in space: The Maxwell equations show that the change of the magnetic field depends on spatial derivatives of the electric field, and the change of the electric field depends on spatial derivatives of the magnetic field; thus, again, the whole set enables predictions for the same whole set.

But what about quantum mechanics?

Quantum mechanical predictions are similar to classical ones: They predict outcomes of measurements. All possible outcomes of one measurement form a Boolean lattice, as in classical physics. But, other than in classical physics, in quantum mechanics there are incompatible observables. That means, if the state of the system can be described by the necessity of a certain outcome (i.e. in attributing the corresponding property—e.g. a certain position—to the system), then there are possible outcomes of other measurements—e.g. of momentum—that cannot be predicted with certainty, but only with certain probabilities: Such observables are called incompatible. "Almost all" pairs of observables of a system are incompatible.

---

[16] Cf. The paper 'The Logic of Quantum Mechanics" of 1936 by G. Birkhoff and J.v.Neuman, *Annals of Mathematics*, 37(1936)(4): 823–843. doi:10.2307/1968621

[17] cf. the quotation from Friebe (2018) on p. 6

So there we are, at the notorious indeterminism of quantum mechanics. Whereas in classical physics we could always suppose that there is *one* Boolean lattice that comprises all possible predictions about a system, in quantum mechanics there are *several* Boolean lattices for different incompatible observables. The question therefore is, how those Boolean lattices combine into a description of the respective system. In the completed quantum mechanics, to be sure, the combination is described as the orthocomplemented lattice of the closed subspaces of Hilbert space. That is known, in principle, empirically. But can we get at understanding that structure without jumping at once to the conclusion "Hilbert space"?

## 10 Are there alternatives to quantum mechanics?

What do we know about that combination lattice, from general consideration? –

1. There are Boolean sublattices, one for every set of compatible observables.
2. It is an orthocomplemented lattice.
3. There are probability functions defined for any Boolean sublattice of the whole lattice.
4. If there is a necessary prediction that defines the present state of the system, it defines probabilities for all predictions.
5. We can compose two independent systems abstractly into one system. The probability of finding a joint result is the product of the two probabilities involved.

Can we conclude from general considerations more specifically what the quantum mechanical lattice is like? – In the completed quantum mechanics the lattice we use is, abstractly speaking, a *complex projective geometry*. This is, in more abstract terms, what is represented by the lattice of the closed subspaces of Hilbert space.

Finding that lattice from plausible assumptions would be the decisive step in the foundation of quantum mechanics. The '*Plato*' database at Stanford[18] says, in referring to this lattice as '$L(\mathbf{H})$': "The point to bear in mind is that, once the quantum-logical skeleton $L(\mathbf{H})$ is in place, the remaining statistical and dynamical apparatus of quantum mechanics is essentially fixed. In this sense, then, quantum mechanics—or, at any rate, its mathematical framework—*reduces to* quantum logic and its attendant probability theory."

Varadarajan says in his seminal book,[19] "For a long time it has been a desire within the community for finding rather comprehensible postulates that imply the structure of quantum mechanics." Varadarajan was one of those scholars who contributed a lot to meet that desire. There has been, though, no mathematical proof so far that Hilbert space is the only possibility for a structure of quantum mechanics. But, up to now, no other lattice has been found that complies with the rules mentioned.

One property that makes this lattice special is its behavior when two systems are composed. When you take one observable of each of the systems, the compound observable is the direct product of the two separate ones. All those direct products form again a lattice of the same type as the component ones, in this case describing the compound system.—In quantum mechanics we describe the state space of the compound system as the tensor product of the two state spaces. This seems to be a very special combination, because it comprises the direct product of any two observables of the two component systems, and they again form an orthocomplemented lattice of the same type as the component ones, though, naturally, its dimension is the product of the dimensions of the components. I suspect that this special structure gives a handle for justifying the structure of quantum mechanics.

---

[18] https://plato.stanford.edu/entries/qt-quantlog/; chapter 1.4
[19] Varadarajan, V. S., Geometry of Quantum Theory, New York (Springer) 1985

Starting from this structure (the orthocomplemented complex projective geometry), we can introduce the usual quantum mechanics of the $\psi$-function etc. as a special representation of that structure, which facilitates calculating measurable results.

Taking quantum mechanics as a lattice of possible predictions, seems to make it easier to understand. Let me take up, in order to try that out, some of the much discussed stumbling blocks within quantum mechanics.

## 11 Stumbling blocks in quantum mechanics

### *Twofold dynamics*

In classical mechanics, dynamics consists of one law for the time development that governs all changes of the respective system. Not so in quantum mechanics: Here, two entirely independent ways of change of state exist. One way is the change according to the Schrödinger equation, the other is the "collapse of the wave function" by a measurement.
The first way of change is quite similar to the change of a field in classical physics, controlled by an field equation. The other (the "collapse") is specifically quantum.[20] That this latter change of state occurs, is a consequence of the quantum mechanical indeterminism: Because of indeterminism it is impossible to predict the outcome of a measurement; there can be no dynamical law that controls that outcome. Any measurement thus implies a surprise for the one who measures, but the state after the measurement is defined by the result of that measurement – hence that jump of the state, the "collapse". Thus, it seems inevitable in an indeterministic theory to have two quite independent dynamics.
There have been and still are attempts at unifying the dynamics of quantum mechanics. But such attempts can only be undertaken on the grounds of a misinterpretation of indeterminism. A truly indeterministic theory must necessarily comprise the two ways of change of state described above.

### *Action at a Distance (EPR)*

The "collapse of the wave function" by a measurement instantly changes the wave function in all of space. This change of state looks like an action at a distance—which is physically impossible. But in the case of an indeterministic theory, a change of the probabilities is nothing but a change of the expectations we have, not a change in the objects we describe; and those can be applicable to as far distant events as one likes: They do not change the reality at any distance!

---

[20] cf. Chapter VI.1 in: Johann von Neumann, Mathematische Grundlagen der Quantenmechanik, Springer, Berlin (1932); English version: John von Neumann, Mathematical Foundations of Quantum Mechanics, Princeton University Press, Princeton, NJ (1955)

A special, more complex case is the conservation of certain quantities in two separate objects—treated usually under the catchword of "EPR". That acronym refers to a paper by Einstein and two of his collaborators, Boris Podolsky and Nathan Rosen, about the question of the "completeness" of quantum mechanics[21]. They start with the definition: "*If, without in any way disturbing a system, we can predict with certainty (i. e. , with probability equal to unity) the value of a physical quantity, then there exists an element of physical reality corresponding to this physical quantity.*" They illustrate their point with an example of measurements of position and momentum; we prefer the example introduced by David Bohm with measurements of the spin components of a spin ½ particle[22]: Imagine a certain particle with spin = 0 that decays into two particles with spin ½ each. Conservation of angular momentum requires that the spins of the two resulting particles are in opposite directions. Thus, I can measure the spin component of one of the particles and infer from the result the spin component of the other, maybe very distant one. From their principle quoted above, EPR conclude that the spin component of the distant particle is an element of reality. The point that makes this conclusion very amazing is that in the quantum mechanical case, the experimenter can decide about the direction of the instrument he measures the spin component with, such that he seems to be able to control an element of reality that might even be light years distant from him. Is that a case of the notorious "spukhafte Fernwirkung" Einstein condemned in a letter to his friend Max Born?[23] I think EPR is another case of the same structure as the "collapse" mentioned above. For an explanation, let us have a look on what really happens in the kind of correlation measurements described.

What happens there in the real world? – We call the two experimenters who measure the spin components on both sides, Alice and Bob, as usual. Suppose Alice measured the spin of her particle with her apparatus oriented vertically and found spin "up". What is the effect of her result on Bob's side? - Bob will see a random sequence of ups and downs (in whichever orientation his apparatus is) with about 50 % of each kind. And that is the important observation: He does not see any effect of what Alice is doing or finding on her side. Actually, there is no "spooky action at a distance", there is no action at a distance, there is even no action at all!

So we have to ask again what it means that the state of the far particle is changed instantly. The answer is analogous to the one given for the "collapse": It means that Bob can construct out of his events a statistical ensemble that represents a certain state *if* he uses the information about Alice's results. If Bob gets the spin result of every single measurement Alice made, then he will find e.g., if his apparatus is oriented vertically as Alice's, comparing Alice's results with his own one by one, that the two results are always opposite. He could filter the sequence of results according to Alice's findings; he could take e.g. all events with spin "up" on Alice's side. In that way he would prepare a new ensemble on his own side corresponding to a

---

state with spin "down". He can do that if he uses the information about the exact sequence of results on Alice's side. That selection really defines a quantum mechanical state: Even if Bob uses his apparatus in a different orientation for the measurements on his side, the frequency of results he finds corresponds to the probability he could have calculated for the quantum mechanical state "down" on his side.

Thus, it is true that the state of the system at Bob's side depends on the results of Alice's measurements. But Bob can find out anything about that state only if he uses the sequence of Alice's results. He must have gotten those results from another source – maybe from an email by Alice. Thus, the change of state by the supposed "spooky action at a distance" is actually not a physical process at all but rather a process of bookkeeping that takes place after the measurements.

This is a way we can understand the rather enigmatic description saying that Bob's state changes instantly but that this change cannot be used to transmit information. Actually nothing changes at Bob's side at all. It is only a later bookkeeping that can give us an opportunity to check the statistical predictions which follow from the state change.

Sometimes it is maintained that the Bell inequalities[24] imply action at a distance in quantum mechanics. But this is not so. What the Bell inequalities say is: In a local hidden variable theory some probability distributions of quantum mechanics cannot be reproduced. If someone believes in hidden variable theories, he is forced to introduce action at a distance. But that depends on his belief in hidden variable theories. I should rather conclude that a hidden variable theory of quantum mechanics is not possible, since locality is of fundamental importance for physics.

There is another issue that has been discussed a lot during those 100 years of quantum mechanics:

## *Theory of Measurement*

There has always been some interest in the quantum mechanical theory of measurement—much more than in the case of classical theories. This is perhaps mainly due to the fact that the things quantum mechanics usually deals with are visible only indirectly; one needs some measurement apparatus in order to know anything about those things. Another, deeper, reason is the fact already mentioned that quantum mechanics does not allow a direct conclusion on the properties of the objects in question but rather only *probabilities* of finding certain results. Thus, the question arises how theoretical results are connected with "real" findings. The theory of quantum mechanical measurement is supposed to provide the bridge between the theory and the findings.

In the course of the development of quantum mechanics there has been so much discussion about measurement that I cannot even give a survey. I will deal shortly with one problem only that is often referred to as the central problem of the theory of measurement. This needs some introductory remarks:

Let us describe the simplest form of a measurement: We start with an observable, say *X*, that is to be measured on a system *S*. The *n* possible results of the measurement are $x_1, \ldots, x_n$.

---

[24] Bell, J. S.: 1964. On the Einstein Podolski Rosen Paradox. Physics l, 195; Reprinted in: Bell, J. S.: 1987. Speakable and Unspeakable in Quantum Mechanics. Cambridge (UP)

For the measurement we use a measuring apparatus $A$ with $n$ possible readings (say, on a scale) $a_1,…, a_n$ corresponding to the $n$ possible results of the measurement. Before the measurement, $S$ and $A$ are separated, $A$ is in some pre-measurement, state. Then the measurement interaction takes place; as a result of that interaction, $A$ shows the reading, say, $a_k$ which means that $S$ is in the state $x_k$.

Now we try describing the interaction in quantum mechanical detail: The originally separated objects $S$ and $A$ interact, i.e. they must be described as *one* compound object, name it *S&A*. This object is transformed according to the Schrödinger equation. In the resulting state there are no more separate objects but there is the compound object *S&A* only, in a new state. Unfortunately this is not what we expected from the measurement; we rather went into all the trouble of measuring in order to end up with a clear cut result $a_i$.

Let us then regard the process of measurement from its expected end: We expect as the result of a measurement one of the readings $a_i$ on the apparatus A. Since we cannot say beforehand which of the $a_i$ will come out, we describe the expected resulting state as a mixture of all possible results, weighted with their probability: $\sum_{1}^{n} p_i x_i$, where $p_i$ is the probability of the result $x_i$. – This second description contains nothing but the state of the system, the apparatus is somehow eliminated. How can we describe the transition of the result of the Schrödinger transformation described above into this last state of the apparatus? – That is the core of the most discussed problem of the quantum mechanical theory of measurement!

What is the status of that strange "mixture"? – Already John von Neumann, in his book of 1932[25], jumps quickly, just by "that is", from the single measurement to the statistical mixture. He writes, after describing the "causal" change of state according to the Schrödinger equation: "On the other hand the state $\varphi$—which may refer to a quantity with a pure discrete spectrum, distinct eigenvalues and eigenfunctions $\varphi_1, \varphi_2 …$— undergoes in a measurement a change in which any of the states $\varphi_1, \varphi_2, …$ may result, and in fact do result with the respective probabilities

$|(\varphi,\varphi_1)|^2$, $|(\varphi,\varphi_2)|^2,…$ . That is the mixture $U' = \sum_{n=1}^{\infty} |(\varphi,\varphi_n)|^2 \cdot P_{[\varphi_n]}$ obtains. […] Since the states go over into mixtures, the process is not causal."[26]

Thus, without further comment, just by "That is", von Neumann jumps directly from the single result of a measurement to the weighted mixture of all possible results. So, for John von Neumann this change is the transition from the state before the measurement to the mixture after the measurement, and, as far as I can see, all later discussions of the process of measurement do the same.

The problem is that the measuring process is described from two different points of view:
1. The dynamics of the measurement interaction results in a pure state of the compound system.

---

[25] von Neumann (1932/1955/2018): Mathematical Foundations of Quantum Mechanics, by John von Neumann. German original 1932, translated from the German 1955 by Robert T. Beyer, New Edition 2018 edited by Nicholas A. Wheeler. Princeton (UP) 2018, p. 417–418
[26] von Neumann John (1955/2018), p 271

2. The description of the expected result of the measurement is a statistical mixture of the possible results, weighted with their probabilities.

The transition from the pure state (1) to the mixture (2) can be described as the "disappearance of the interference terms" of the state. There have been many attempts at finding a quantum mechanical process that would accomplish that change. Peter Mittelstaedt, e.g., spent most of his academic life with such attempts. Towards the end of his life he seems to have given up hope of finding a solution. – Even a rather new account of the measurement problem[27] gives exactly the the same representation of the problem as John von Neumann in 1932.

My impression is, on the other hand, that such a process is not necessary[28]. It is rather that you have to make up your mind on what you are talking about:

One might talk about a beam of particles passing through a device that sorts the particles according to the eigenvalues of the observable to be measured. There we arrive at a mixture, either of a collection of empirical results with there frequencies or, within the theory, a collection of all possible results with the respective probabilities: it is a formal collection of all (real or possible) results of the measurement into a statistical ensemble. But this ensemble is the result of a bookkeeping process. It is an ensemble of 'classical' results; quantum mechanics must not be applied to them, so there can be no question of interference terms from the beginning.

On the the other hand one might talk about a single measurement. In one real experiment there is only one result (albeit unpredictable, as a consequence of indeterminism). Thus, again, there is no question of interference terms.

## *Realism*

Realism in quantum mechanics is a subject that has been intensely discussed. This is quite understandable since one of the characteristic features of quantum mechanics is that it does not give a picture of reality, like classical physics does. It gives rules for predictions on measurements instead. So, many physicists tried to give quantum mechanics a "realistic" interpretation, i.e. an interpretation that describes a reality "out there" that is independent of being measured or even perceived at all. Advocates of a realistic view seem to consider it self-evident that there must be some reality "behind" the phenomena described by quantum mechanics[29]. There is, e.g., a paper by Tim Maudlin[30] where the author tries to analyze the problem of quantum mechanical measurement. In the beginning he states three inconsistent claims. I quote:

"The following three claims are mutually inconsistent:

1.A The wave-function of a system is complete, i.e. the wave-function specifies (directly or indirectly) all of the physical properties of a system.

---

[27] Holger Lyre. Why Quantum Theory is Possibly Wrong. Found Phys **40**(2010)1429–1438
[28] Michael Drieschner, A Note on the Quantum Mechanical Measurement Process. philosophia naturalis **50**(2013)201–213
[29] cf. p. 6 above
[30] Tim Maudlin, Three Measurement Problems. Topoi **14**(1995)7-15

> 1.B The wave-function always evolves in accord with a linear dynamical equation (e.g. the Schrödinger equation).
>
> 1.C Measurements of, e.g., the spin of an electron always (or at least usually) have determinate outcomes, i.e., at the end of the measurement the measuring device is either in a state which indicates spin up (and not down) or spin down (and not up)."

It is already statement 1.A that does not really fit quantum mechanics: The wave function never specifies physical properties of a system; this would presuppose some kind of "classical" understanding. What the wave function (or, more generally, the quantum mechanical state) specifies is a catalog of probabilities for all possible results of experiments on the system. Thus, 1.A can never be true, independently of the other claims.
1.B is not quantum mechanical either: As quoted above, John von Neumann stated already in his (1932) that there are two ways the quantum mechanical state changes in time according to quantum mechanics, namely either according to the Schrödinger equation or by a "collapse" after a measurement. – cf. the discussion above.

We see that Maudlin's text is biased in favor of some "realistic" world view that is not applicable in quantum mechanics.

## .1 Bohm

A well-known example of realism, and probably the oldest one, is David Bohm's theory, first published in 1952[31]. Bohm says there that he reformulated quantum mechanics as a basis for his attempt at extending quantum mechanics in order to get a "realistic" theory. His reformulation suggested a similarity with Hamilton-Jacoby theories of classical mechanics. But in fact it was nothing but quantum mechanics in a formalism a bit unusual. In those about 70 years since, nobody succeeded really extending quantum mechanics in a way Bohm probably had in mind. But "Bohmian Mechanics" adds to quantum mechanics a way of speaking about nature that were not possible if one regarded nothing but measurable quantities and feasible experiments. In Bohmian Mechanics there are, e.g., trajectories of particles; trajectories do not exist in quantum mechanics. Even according to "Bohmians" those trajectories are principally not visible or measurable. In fact, they are nothing but words, added to quantum mechanics. Besides, particles, according to the Bohmian way of talking, would have speeds faster than light, sometimes they would even have infinite velocity. Bohmians say that this does not matter because they can never be observed anyway.

There is an excellent thorough representation of Bohmian Mechanics by a true Bohmian, Oliver Passon.[32]

---

[31] Bohm, D.: 'A Suggested Interpretation of the Quantum Theory in Terms of "Hidden" Variables.' Phys. Rev. **85**(1952)166,180; reprinted in: Wheeler, J.A. und Zurek, W.H. (eds.): Quantum Theory and Measurement. Princeton, NJ (UP) 1983

[32] Passon, Oliver. Bohmsche Mechanik: Eine elementare Einführung in die deterministische Interpretation der Quantenmechanik. Harri Deutsch, Frankfurt, 2004, ISBN 978-3-8171-1742-0. – There exists no English translation.

### *.2 Many Worlds*

An other attempt at "improving" quantum mechanics, which is as well known as Bohmian Mechanics, is the "Many Worlds" interpretation by Everett, Wheeler, and DeWitt[33]. I must confess that I do not understand why this interpretation shows up in many discussions as an alternative to the usual interpretation of quantum mechanics. It is again, looked at closely enough, neither an alternative theory nor an alternative interpretation of quantum mechanics. It is, basically, nothing but a change of language: The usual description of quantum mechanics talks of a set of possibilities (for a measurement outcome), one of which becomes real in the measurement, the others don't. What is usually called the Everett-Interpretation says, on the other hand, that *all* possibilities become real. In order to make that possible, the Everettism says that the universe splits up into as many universes as there are possible outcomes, and every possible outcome becomes real in one of those universes. What does that mean? – I cannot find any rational interpretation of that story. Let us try to translate the Everettish language back to ordinary English: Instead of talking of "other universes" one could just as well talk about "possibilities not realized"—this would not change anything at all. Thus, what remains of the "Everett-interpretation" is only an unusual wording that is easily re-translatable, though, into ordinary language.

## 12 Conclusion

It is not so hard to understand quantum mechanics, once one has accepted its indeterministic character (and uses the lattice representation). Understanding the underlying concept of probability is much easier if one starts from the definition: *probability is predicted relative frequency.* Thus, Fuchs and Peres couldn't be more right: *Quantum Theory needs no 'interpretation'*!

∎

---

[33] Everett, H.: 1957. "Relative State" Formulation of Quantum Mechanics. Rev. Mod. Phys. 29, 454; Wheeler, J.A.:1957: Assessment of Everett's "Relative State" Formulation of Quantum Theory. Rev. Mod. Phys. 29, 463; DeWitt, Bryce: 1970: Quantum mechanics and Reality. Physics Today Sept. 1970, 30-35.

Quotation from: Friebe et al 2018, p. 39

If one tries to proceed systematically, then it is expedient to begin with an interpretation upon which everyone can agree, that is with an instrumentalist minimal interpretation. In such an interpretation, Hermitian operators represent macroscopic measurement apparatus, and their eigenvalues indicate the measurement outcomes (pointer positions) which can be observed, while inner products give the probabilities of obtaining particular measured values. With such a formulation, quantum mechanics remains stuck in the macroscopic world and avoids any sort of ontological statement about the (microscopic) quantum-physical system itself.

*(Friebe (2018), p.44)*

## 2.2 The Ensemble Interpretation and the Copenhagen Interpretation

The first stage of interpretation of the mathematical formalism establishes the connection to the empirical world as far as needed for everyday physics in the laboratory or at the particle collider. Born's rule allows a precise prediction of the probabilities of observing particular outcomes in real, macroscopic measurements. The fact that this minimal interpretation makes statements only about macroscopic, empirically directly accessible entities such as measurement setups, particle tracks in detectors or pulses from a microchannel plate may be quite adequate for those who see the goal of the theory within an experimental science such as physics as being simply the ability to provide empirically testable predictions. For the metaphysics of science, this is not sufficient, and most physicists would also prefer to have some idea of what is behind those measurements and observational data, i.e. just how the microscopic world which produces such effects is really structured. In contrast to the instrumentalist minimal interpretation, however, every additional assumption which might lead to a further-reaching interpretation remains controversial.